\documentclass[]{bmcart}
\usepackage{amsthm,amsmath}
\usepackage{bm}
\usepackage[utf8]{inputenc} 
\usepackage{graphicx}
\graphicspath{{./Figs/}}
\usepackage{cancel}
\usepackage{booktabs}

\usepackage{upgreek}
\usepackage{graphicx}
\usepackage{subfigure}
\usepackage{color}
\usepackage{color}
\usepackage[normalem]{ulem} 
\newcommand{\Replace}[2]{\bgroup\noindent\textcolor{red}{\xout{#1} #2}\egroup\ignorespacesafterend}
\newcommand{\Delete} [1]{\bgroup\noindent\textcolor{red}{\xout{#1}}\egroup\ignorespacesafterend}
\newcommand{\Insert} [1]{\bgroup\noindent\textcolor{}{#1}\egroup\ignorespacesafterend}
\newcommand{\Comment}[1]{\definecolor{Mygray}{gray}{0.50}\bgroup\color{Mygray}\noindent#1\egroup\ignorespacesafterend}
\newcommand \Michael [1]{\bgroup\noindent[\textcolor{blue}{\textbf{Michael}: #1}]\egroup\ignorespacesafterend}
\newcommand \Stefan  [1]{\bgroup\noindent[\textcolor{blue}{\textbf{Stefan}: #1}]\egroup\ignorespacesafterend}

\DeclareMathAlphabet{\Ibb}{U}{msb}{m}{n}

 \newcommand{\dx}{\text{\rm d}x}
 \newcommand{\dy}{\text{\rm d}y}
 
 \newcommand{\ds}{\text{\rm d}s}

\newcommand{\taue}{\tau_{\rm ext}}

\newcommand{\Bb}{{\boldsymbol{\mathnormal b}}}

\newcommand{\Be}{{\boldsymbol{\mathnormal e}}}

\newcommand{\Br}{{\boldsymbol{\mathnormal r}}}
\newcommand{\Bs}{{\pmb{\mathnormal s}}}

\newcommand{\erf}{\textrm{erf}}

\newcommand{\figref}[1]{Fig.~\ref{#1}}

\newcommand \MZ [1] {\bgroup\noindent[\textcolor{blue}{\textbf{MZ}: #1}]\egroup\ignorespacesafterend}

\begin{document}

\begin{frontmatter}

\begin{fmbox}
\dochead{Research}

\title{Pinning of extended dislocations in atomically disordered crystals}

\author[addressref={aff1},            
]{\inits{AV}\fnm{Aviral} \snm{Vaid}}
\author[addressref={aff2},                   
]{\inits{DW}\fnm{De'an} \snm{Wei}}
\author[addressref={aff1},                
]{\inits{EB}\fnm{Erik} \snm{Bitzek}}
\author[addressref={aff3},                 
]{\inits{SN}\fnm{Samaneh} \snm{Nasiri}}
\author[addressref={aff3},corref={aff3},email={michael.Zaiser@fau.de}
]{\inits{MZ}\fnm{Michael} \snm{Zaiser}}

\address[id=aff1]{
\orgname{Department of Materials Science, WW1-General Materials Properties, Friedrich-Alexander Universität Erlangen-Nürnberg}, 
  \street{Martensstr. 5},                     %
  \postcode{91058}                                
  \city{Erlangen},                              
  \cny{Germany}                                    
}
\address[id=aff2]{
	\orgname{Applied Mechanics and Structure Safety Key Laboratory of Sichuan Province, School of Mechanics and Engineering, Southwest Jiaotong University},                    %
	\postcode{610031}                                
	\city{Chengdu},                              
	\cny{China}                                    
}
\address[id=aff3]{
\orgname{Department of Materials Science, WW8-Materials Simulation, Friedrich-Alexander Universität Erlangen-Nürnberg}, 
  \street{Dr.-Mack-str. 77},                     %
  \postcode{90762}                                
  \city{Fürth},                              
  \cny{Germany}                                    
}

\begin{artnotes}

\end{artnotes}

\end{fmbox}


\begin{abstractbox}

\begin{abstract} 
In recent years there has been renewed interest in the behavior of dislocations in crystals that exhibit strong atomic scale disorder, as typical of compositionally complex single phase alloys. The behavior of dislocations in such crystals has been often studied in the framework of elastic manifold pinning in disordered systems. Here we discuss modifications of this framework that may need to be adapted when dealing with extended dislocations that split into widely separated partials. We demonstrate that the presence of a stacking fault gives rise to an additional stress scale that needs to be compared with the pinning stress of elastic manifold theory to decide whether the partials are pinned individually or the dislocation is pinned as a whole. For the case of weakly interacting partial dislocations, we demonstrate the existence of multiple metastable states at stresses below the depinning threshold and analyze the stress evolution of the stacking fault width during loading. In addition we investigate how geometrical constraints can modulate the dislocation-solute interaction and enhance the pinning stress. We compare our theoretical arguments with results of atomistic and discrete (partial) dislocation dynamics (D(P)DD) simulations. 
\end{abstract}

\begin{keyword}
\kwd{High entropy alloys}
\kwd{partial dislocations}
\kwd{stacking fault}
\kwd{critical resolved shear stress}
\kwd{atomistic simulation}
\kwd{discrete dislocation dynamics simulation}
\end{keyword}

\end{abstractbox}

\end{frontmatter}
´´
\section[Introduction]{Introduction}

The theory of dislocations interacting with atomic-scale obstacles, traditionally formulated in the context of solution hardening, has seen a renaissance in recent years which has been driven by the general interest in compositionally complex alloy systems including so-called high-entropy alloys. In such alloys, multiple atomic species are present in comparable concentrations and entropic effects may stabilize homogeneous solid solutions  at elevated temperatures, whereas kinetic effects (slow diffusion due to multiple barriers and traps) may stabilize such phases against unmixing at reduced temperature. 

From a theoretical viewpoint, statistical theories of dislocation pinning by atomic-scale obstacles have, starting from the seminal work of Labusch \cite{Labusch1970_PSS,Labusch1972_AM}, attracted the interest of statistical physicists, and concepts developed for the pinning of elastic manifolds by random fields (e.g. \cite{Chauve2000_PRB}) and their depinning by external forces were applied to the athermal motion of dislocations (e.g. \cite{Zapperi2001_MSEA,Bako2008_PRB}) and to dislocation motion at finite temperatures \cite{Ioffe1987_JPC, Zaiser2002_PM}. In recent years, these concepts have been extended and adapted to compositionally complex and high entropy alloys by a number of authors (e.g.\cite{Toda2015_AM,Wu2016_AM,Varvenne2016_AM,Varvenne2017_AM,larosa2019solid,zhai2019properties,zaiser2021pinning}). 
One aspect which has received little attention in the statistical mechanics literature and in classical theories of solute hardening is the fact that dislocations split into partials, which provides them with an additional degree of freedom (the width of the stacking fault separating the partials) when interacting with localized obstacles. This aspect may be of particular importance in
concentrated solid solution alloys, which are also referred to as compositionally complex or high entropy alloys,  where stacking fault energies may be quite low and, accordingly, partial dislocations may be widely separated while at the same time the resistance to their motion is high due to the dislocation-solute interactions \cite{shih2021stacking}. In the present paper we study the consequences of core splitting on dislocation interactions with atomic scale disorder. In Section 2, we briefly review the scaling ideas of elastic manifold (de)pinning for the case of dislocations treated in a line tension approximation. In Section 3, we discuss the modifications to this picture that arise when dislocations are split into partials and derive some predictions, which we compare in Section 4 with results of atomistic simulations of dislocations in compositionally complex alloys. In the concluding Section 4, we discuss implications of our findings.

\section{Scaling theory of elastic lines in static random fields}
\label{sec:2} 

We envisage the dislocation as an elastic line of line tension ${\cal T}= C_{\rm LT}\mu b^2$ where $\mu$ is the shear modulus, $b$ the length of the Burgers vector, and the numerical parameter $C_{\rm LT}$ depends on the dislocation character, as well as logarithmically on geometrical parameters characterizing the line shape. The dislocation is assumed macroscopically straight while local fluctuations of the line shape are described by a function $y(x)$ where the $x$ axis is oriented along the average line direction and the $y$ axis in perpendicular direction within the glide plane. Note that we assume $y(x)$ to be single valued, which excludes the presence of overhangs, which are indeed absent in our atomistic simulations of dislocations in compositionally complex alloys. 

Let atomic disorder create a spatially fluctuating but temporally fixed resolved shear stress field $\tau(x,y)$ that acts on the dislocation, which is in addition subject to a spatially constant resolved shear stress $\taue$. The evolution of the dislocation line shape is then given by the quenched Edwards-Wilkinson equation 
\begin{equation}
	\frac{b}{B} \frac{d y}{d t} = {\cal T}\frac{\partial^2 y}{\partial x^2} + b [\tau(x,y) + \taue],
	\label{eq:qedwilk}
\end{equation}
with the damping or drag coefficient $B$. 
 Owing to the $y$ dependence of the random stress, Eq. \eqref{eq:qedwilk} is intrinsically nonlinear. The random shear stress $\tau(x,y)$ has a correlation function of the form
\begin{equation}
	\langle \tau(x,y)\tau(x',y') \rangle = \langle \tau^2 \rangle  \Psi(x-x',y-y')
\end{equation}
where the magnitude $\langle \tau^2 \rangle$ can for a random alloy be related to the average of the squared atomic misfit \cite{geslin2021microelasticityI}. The functions $\Psi_{\parallel}(x) := \Psi(x,0)$ and $\Psi_{\perp}:= \Psi(0,y)$ where $\Psi_{\parallel}(0) =  \Psi_{\perp}(0)=1$ describe correlations of the random shear stress field in the directions parallel and perpendicular to the dislocation line, respectively. We define the correlation length of the pinning field as $\xi = \int \Psi_{\parallel} \dx$, and it is understood that we consider the line on scales well above the correlation length such that the functional form of the correlations in the line-parallel direction does not matter. Explicit expressions for the correlation functions that arise from atomic misfit in a random alloy can be found in Ref.\cite{geslin2021microelasticityII}, we give these expressions and the related integrals in the appendix.

The energy landscape of the dislocation is governed by the interplay of the fluctuating stress field -- which wants to displace the dislocation according to the sign of the locally acting shear stress -- and the line tension ${\cal T} \propto \mu b^2$, which penalizes rapid fluctuations. The ensuing mean energy change for a segment of  length $L \gg \xi$ displacing coherently over a distance $w$ is given by 
\begin{equation}
\Delta E(L,w) = E_{\rm LT} - b \left\langle \int_0^w \tau(x,y) \dx \right\rangle_L
\label{eq:elw}
\end{equation}
Here the average $\langle \dots \rangle_L$ denotes an average over the interval $y \in [-L/2\dots L/2]$. The evaluation of this average is detailed in Appendix C. Furthermore, we estimate the line tension energy (per unit length) associated with a displaced bulge of length $L$ and depth $w$ as $E_{\rm LT} = C_{\rm LT} \mu b^2 (w^2/L^2)$ where the pre-factor $C_{\rm LT}$ depends on dislocation orientation as well as (weakly) on geometrical parameters of the displaced segment. The result is
\begin{equation}
	\Delta E(L,w) = C_{\rm LT}\mu b^2\frac{w^2}{L^2} - b \sqrt{\langle \tau^2 \rangle \frac{\xi}{L}}\int_0^w \Psi_{\perp}(x) \dx 
	\label{eq:elw}
\end{equation}
The pinning energy per unit length derives by minimizing Eq. (\ref{eq:elw}) with respect to $L$ and $w$. Setting $\partial_w \Delta E(L,w) = \partial_L \Delta E(L,w) = 0$ gives 
\begin{eqnarray}
	-2 C_{\rm LT} \mu b^2 \frac{w_{\rm p}^2}{L_{\rm p}^3} + \frac{b}{2} \sqrt{\left\langle \tau^2 \right\rangle \frac{\xi}{L_{\rm p}^3}} \int_0^{w_{\rm p}} \Psi_{\perp} \dy &=& 0
	\nonumber\\
	2 C_{\rm LT} \mu b^2 \frac{w_{\rm p}}{L_{\rm p}^2} - b \sqrt{\left\langle \tau^2 \right\rangle \frac{\xi}{L_{\rm p}}} \Psi_{\perp}(w_{\rm p}) &=& 0
\end{eqnarray}
It follows that the optimal displacement depends only on properties of the perpendicular correlation function as it solves the equation
\begin{equation}
2 w_{\rm p}\Psi_{\perp}(w_{\rm p}) = \int_0^{w_{\rm p}} \Psi_{\perp}(y)\dy .
\end{equation}
The pinning length follows as 
\begin{equation}
	L_{\rm p} = \left(\frac{2C_{\rm LT}}{\Psi_{\perp}(w_{\rm p})}\right)^{2/3}\left(\frac{\mu}{\sqrt{\langle\tau^2\rangle}}
	\right)^{2/3} \left(\frac{b^2 w_{\rm p}^2}{\xi}\right)^{1/3}
	\label{eq:lpin1}
\end{equation}
Inserting (5) and (6) into (3) gives the pinning energy which results as:
\begin{equation}
	E_{\rm p} = -3\left(\frac{\Psi_{\perp}(w_{\rm p})^4}{16C_{\rm LT}} \right)^{1/3}\frac{\sqrt{\langle \tau^2 \rangle}^{4/3}}{\mu^{1/3}} (\xi b w_{\rm p})^{2/3} = C_E(\theta)\frac{\sqrt{\langle \tau^2 \rangle}^{4/3}}{\mu^{1/3}}b^2
	\label{eq:epin}
\end{equation}
The critical shear stress can be estimated by equating the pinning energy to the work done by a spatially uniform stress in displacing the dislocation over the distance $w_{\rm p}$:
\begin{equation}
	\tau_{\rm c} = \frac{E_{\rm p}}{w_{\rm p}b} = C_{\tau}(\theta)\frac{\sqrt{\langle \tau^2 \rangle}^{4/3}}{\mu^{1/3}}.
	\label{eq:taupin}
\end{equation}

\section{Scaling theory of a split dislocation with weakly interacting partials}

A split dislocation is made up of two partials separated by a stacking fault of energy $\Gamma_{\rm SF}$. In the following variables referring to the leading and trailing partial are labeled by the subscripts 1 and 2, respectively. The Burgers vectors of the two partials are $\Bb_{\rm p1}$ and $\Bb_{\rm p2}$. Both partial Burgers vectors have equal length $b_{\rm p}$ and are contained in the slip plane of the original dislocation of Burgers vector $\Bb$. The Burgers vectors fulfill the relations
\begin{equation}
	\Bb_{\rm p1}+\Bb_{\rm p2}=\Bb\quad,\quad
	\Bb_{\rm p1}^2+\Bb_{\rm p2}^2 < b^2 \quad,\quad
	\Bb_{\rm p1}.\Bb = \Bb_{\rm p2}.\Bb = b^2/4\quad.
\end{equation} 
The equilibrium distance of the two partials is 
\begin{equation}
	d_0 = \frac{C_{\Gamma}\mu b_{\rm p}^2}{\Gamma_{\rm SF}}
\end{equation}
and the constant $C_{\Gamma}$ depends on the dislocation character according to \cite{hirsch1965stacking}
\begin{equation}
	C_{\Gamma} = \frac{1}{8\pi}\frac{2-\nu}{1-\nu}\left(1 - \frac{2\nu \cos2\theta}{2-\nu}\right)
\end{equation}
where $\theta$ is the angle between line direction and Burgers vector of the full dislocation. 
The dislocation is then described by two coupled equations of motion for the two partials with line shapes $y_1(x_1)$  and $y_2(x_2)$ where it is understood that $y_1 > y_2$. The equations of motion read
\begin{eqnarray}
	\frac{b}{B} \frac{d y_1}{d t} &=& C_{\rm LT,1} \mu b_{\rm p}^2\frac{\partial^2 y_1}{\partial x_1^2} + \frac{b}{2}  \taue + b_{\rm p} \tau(x_1,y_1) +f_{1}(y_1,y_2)\\
	\frac{b}{B} \frac{d y_2}{d t} &=& C_{\rm LT,2} \mu b_{\rm p}^2\frac{\partial^2 y_2}{\partial x_2^2} + \frac{b}{2} \taue  + b_{\rm p} \tau(x_2,y_2) + f_{2}(y_1,y_2).
	\label{eq:qedwilk2}
\end{eqnarray}
The interaction forces $f_{1/2}(y_1,y_2)$ are computed in the Appendix. We consider the limiting case where the interaction between the partials is weak, such that the force exerted by the stacking-fault is less than the pinning force on each partial due to its interaction with the fluctuating shear stresses due to atomic misfit. This is possible in some high-entropy alloys where stacking fault energies may be close to zero. In this case, the interactions between fluctuations of both partials can be neglected, and we can treat the interaction between the partials in a {\em mean-field approximation} where fluctuations on each partial interact only with the mean position of the second partial. The interaction force then depends in leading order only on the average stacking fault width and we can write
\begin{eqnarray}
	\frac{b}{B} \frac{d y_1}{d t} &=& C_{\rm LT,1} \mu b_{\rm p}^2\frac{\partial^2 y_1}{\partial x_1^2} + b_{\rm p}\tau_1(x_1,y_1) + \frac{b}{2} \taue + C_{\Gamma}
	\mu b_{\rm p}^2 \left(\frac{1}{d}-\frac{1}{d_0}\right)\\
	\frac{b}{B} \frac{d y_2}{d t} &=& C_{\rm LT,2}\mu b_{\rm p}^2\frac{\partial^2 y_2}{\partial x_2^2} + b_{\rm p} \tau_2(x_2,y_2) +\frac{b}{2}  \taue - C_{\Gamma}
\mu b_{\rm p}^2 \left(\frac{1}{d}-\frac{1}{d_0}\right).
\end{eqnarray}
Note that, for a partial dislocation, we need to distinguish the shear stress in the 'partial slip system' with slip vector $\Bs_{\rm p} = \Bb_{\rm p}/b_{\rm p}$ and the shear stress in the slip system of the full dislocation, with slip vector $\Bs = \Bb/b$. The fluctuating stress exerted by a random arrangement of atoms of varying size, which leads to pinning, must be considered for the two partials in the respective partial slip system. The external shear stress which drives both partials, on the other hand, is defined in the slip system of the full dislocation. When we add Peach-Koehler forces arising from stresses of both types, we always refer to the slip system of the full dislocation; this implies that the pinning stress in this case is multiplied with a factor $2b_{\rm p}/b$ (see full and dashed green lines in Figure 1, right)

These equations can be envisaged as describing the motion of two near-independent dislocations of Burgers vector $b_{\rm p}$ in effective stress fields 
\begin{equation}
	\tau_{\rm eff,1} = \taue + C_{\Gamma}
	\mu b_{\rm p} \left(\frac{1}{d}-\frac{1}{d_0}\right)\quad,\quad
	\tau_{\rm eff,2} = \taue - C_{\Gamma}
	\mu b_{\rm p} \left(\frac{1}{d}-\frac{1}{d_0}\right).
\end{equation}
Note that, for weakly interacting partials, these stress fields are spatially near-uniform on the scale of the partial dislocation shape fluctuations that lead to pinning. 

To each partial dislocation we can thus separately apply the scaling considerations of the previous section. As compared to the perfect dislocation, the shorter Burgers vector implies a reduced line tension which according to Eq. (7) leads to a reduced pinning length and, according to Eqs. (8,9), an increased pinning stress. Moreover, the partial dislocation character and as a consequence the pinning parameters are different from those of the perfect dislocation. We illustrate these differences in Section 4.1.

\section{Results}

\subsection{Pinning parameters for extended dislocations}

\begin{figure}[tb]
	\centering
	\hbox{}
	\includegraphics[width=0.7\textwidth]{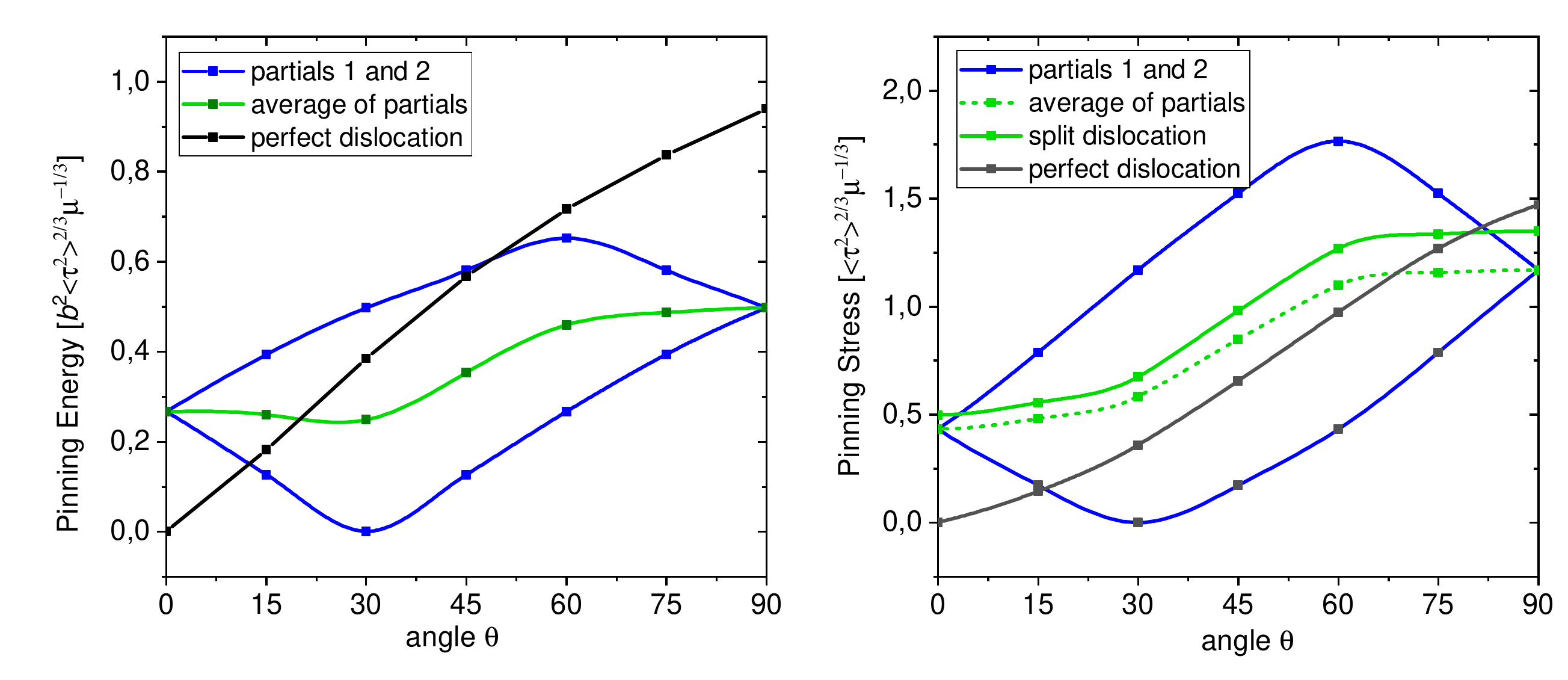}\hfill
	\caption{\label{fig:scaling}
		Pinning energy and pinning stress for partial dislocations in a fcc lattice, limit of weak coupling. Parameters are given as function of the angle between Burgers vector and line direction of the full dislocation. Parameters for a perfect dislocation are given for comparison. Left: Pinning energy, right: pinning stress. Blue lines indicate parameters for partial dislocations, green lines averages of partial dislocations, black lines parameters for the perfect dislocation. In the right graph the dashed green line gives the average of the pinning stresses in the partial slip systems, and the full green line the overall pinning stress of the extended dislocation.}
\end{figure}

We evaluate, as a function of dislocation orientation, the pinning parameters for extended dislocations. We first consider the case of weakly interacting partials where we can consider the partials separately. The pinning energy and depinning stress depend on dislocation orientation through several factors, namely (1) the line tension parameter $C_{\rm LT}$ and (2) the pinning displacement $w_{\rm p}$ and associated correlation function value $\Phi_{\perp}(w_{\rm p})$. The latter dependency arises because the correlations of the shear stress field are anisotropic with respect to the Burgers vector direction of a dislocation \cite{geslin2021microelasticityII}.

Using the correlation functions of Ref. \cite{geslin2021microelasticityII}, orientation dependent pinning parameters have been evaluated in Ref. \cite{zaiser2021pinning}. Here we use these results to evaluate the orientation dependent pinning parameters $C_E, C_{\tau}$ of Eqs. (8) and (9) for the perfect dislocation and the corresponding partials. For details of the calculation the reader is referred to \cite{zaiser2021pinning}. Note that for the partial dislocation the depinning stress is understood as the effective stress according to Eq. (17). 

For dislocations of pure edge and screw orientation, both partial dislocations have the same character (30 degree dislocations for screws, 60 degree dislocations for edges) and therefore the same pinning stress. This is not true for dislocations of generic orientation, for which the pinning stresses differ. As a consequence, the partial with the lower pinning stress may start to move at a lower applied stress, until the ensuing change in stacking fault width leads to a situation where the effective stresses on each partial match the respective pinning stresses such that the dislocation can move coherently. We now explore the consequences of this behavior.

\subsection{History and hysteresis effects in depinning}

We again consider weakly interacting partial dislocations where Eq. (17) holds. Static or metastable situations are characterized by the condition that the effective stresses fall below the respective pinning stresses:
\begin{equation}
	\tau_{\rm 1,c} \ge \left|\taue + C_{\Gamma}
	\mu b_{\rm p} \left(\frac{1}{d}-\frac{1}{d_0}\right)\right|\quad,\quad
	\tau_{\rm 2,c} \ge \left|\taue - C_{\Gamma}
	\mu b_{\rm p} \left(\frac{1}{d}-\frac{1}{d_0}\right)\right|\quad,\quad
\end{equation}
Here the '$<$' relation indicates that the respective partial is in a metastable state, and the '=' relation indicates the transition to the moving state. Depending on the magnitude of the external shear stress, the magnitude of the pinning stresses and the stacking fault width $d$ we find a wide variety of combinatorial possibilities and a correspondingly rich behavior. 

\subsubsection{Metastable states at zero external stress}

In the absence of an external stress, metastability is defined by the requirement that the net interaction of both partials falls below the pinning stress of the partial that experiences the weaker pinning:
\begin{equation}
\min(\tau_{\rm 1,c},\tau_{\rm 2,c}) \ge \left| \tau_{\Gamma}  \left(\frac{d_0}{d}-1\right)\right|,
\end{equation}
where we have introduced the notation $\tau_{\Gamma} = 2\Gamma/b = C_{\Gamma}\mu b_{\rm p}/d_0$. Denoting $\min(\tau_{\rm 1,c},\tau_{\rm 2,c})=: \tau_{\rm min}$, we find for $\tau_{\Gamma}>0$ (positive SF energy) the inequality
\begin{equation}
\frac{1}{\tau_{\rm min}/\tau_{\Gamma} +1} \le \frac{d}{d_0} \le	\frac{1}{\tau_{\rm min}/\tau_{\Gamma} -1}
\end{equation}
which defines a band of metastable initial widths. If a dislocation is created, e.g. in an atomistic simulation, with a $d$ value outside this regime, the SF will expand or constrict during relaxation as the dislocation adjusts to the pinning landscape, until the $d$ value reaches the boundary of the metastable band. If the dislocation is created with a $d$ value within the metastable band, on the other hand, the SF width is expected to remain approximately unchanged during relaxation. 

We note that metastable solutions even exist for
$\tau_{\Gamma} < 0$ and $\tau_{\rm min} < |\tau_{\Gamma}|$, when the metastable regime is given by
\begin{equation}
	\frac{1}{\tau_{\rm min}/|\tau_{\Gamma}| - 1} \le \frac{d}{|d_0|} 
\end{equation}
Partial dislocation -- solute interactions may thus prevent complete dissociation of a dislocation even if the SF energy is negative.

\subsubsection{Metastable states and depinning under stress}

We first consider the situation where both partials share the same pinning stress $\tau_{\rm c} = \tau_{\rm 1,c} = \tau_{\rm 2,c}$  (edge or screw orientation of the full dislocation). In this case, coherent motion of both partials dislocations occurs with $d=d_0$ at stresses
	\begin{equation}
		\taue > \tau_{\rm c}. 
	\end{equation}
If the stacking fault width is initially above $d_0$, the trailing partial experiences a higher effective stress than we leading one, whereas the opposite is true for stacking fault width below $d_0$. We thus find from Eq. (21) that the metastable band is defined by
	\begin{equation}
		\frac{d}{d_0}  \ge \frac{\tau_{\Gamma}}{\tau_{\rm c}- \taue +\tau_{\Gamma}}
		\;,\; d < d_0\;;\quad
		\frac{d}{d_0}  \le \frac{\tau_{\Gamma}}{\taue-\tau_{\rm c} + \tau_{\Gamma}}
		\;,\; d > d_0\;.\;
	\end{equation}
Thus, the critical SF width increases with applied stress for $d<d_0$, and it decreases for $d>d_0$. At the critical stress $\tau_{\rm c}$, upper and lower critical branches merge at $d=d_0$. Beyond this point, no metastable states exist.

In the case of dislocations of generic orientation, the pinning stresses of both partials are different. In this case, coherent motion of the dislocation requires the SF width to adjust such that the interaction of both dislocations compensates the pinning stress difference. We assume without loss of generality that a positive external stress is applied and distinguish two situations, depending whether the leading or the trailing partial has the higher flow stress. We find that coherent motion occurs at a stress $\taue = (\tau_{\rm 1,c} + \tau_{\rm 2,c})/2$ once the SF width has adjusted to the dynamic equilibrium value 
	\begin{equation}
		\frac{d}{d_0}  \le \frac{\tau_{\Gamma}}{\Delta\tau_{\rm c}+\tau_{\Gamma}},		
	\end{equation}
where $\Delta \tau_{\rm c} = \tau_{\rm 1,c}-\tau_{\rm 2,c}$. If the leading partial has the higher flow stress ($\Delta \tau_{\rm c}>0$) the SF constricts and the dislocation moves coherently. In the opposite case, when the SF expands, we find a sigularity when $(\tau_{\rm 2,c} - \tau_{\rm 1,c}) > \tau_{\Gamma}$, i.e., when the maximum force that can be exerted by the SF is insufficient to compensate the flow stress difference. As a consequence, the dislocation disintegrates as the leading partial detaches at the stress $\taue = \tau_{\rm 1,c} + \tau_{\Gamma}$ and then moves indefinitely, while the trailing partial remains pinned. Note that stress reversal makes the leading into the trailing partial, such that a dislocation which is stable under positive stress may disintegrate  upon stress reversal and vice versa. A further level of complication is added by the fact that the pinning stresses decrease from edge to screw orientation. Hence, screw dislocations are less prone to disintegration than edges, and curved dislocations may disintegrate partially. 

When the leading partial has a lower pinning stress than the trailing partial, the dislocation may disintegrate despite a positive SF energy. Conversely, if the trailing partial has a lower pinning stress than the leading partial, the dislocation may be able to move coherently even if the SF energy is negative, provided that $\Delta \tau_{\rm c} > |\tau_{\Gamma}|$. The corresponding SF width is
\begin{equation}
	\frac{d}{|d_0|}  = \frac{|\tau_{\Gamma}|}{\Delta\tau_{\rm c}-|\tau_{\Gamma}|},		
\end{equation}
We thus encounter a very wide range of possibilities which, evidently, makes it extremely difficult to infer the SF energy from observations of the SF width of either static or moving dislocations. 

\subsection{Pinning of constrained dislocations}

If the length of dislocation that can move freely is constrained, this imposes an upper limit on the length $L$. As a consequence, the pinning parameters change. In this case, the displacement $w_L$ which optimizes the pinning energy under the constraint of a given $L$ follows from the second Eq. (5) which we re-write using Eq. (7) as
\begin{equation}
	w_L = 	w_{\rm p}\frac{\Phi_{\perp}(w_L)}{\Phi_{\perp}(w_{\rm p})}\left(\frac{L}{L_{\rm p}}\right)^{3/2}.
\end{equation}
For stress fluctuations induced by a random distribution of compression and/or dilatation centers, the correlation functions are near-universal and depend only on dislocation orientation \cite{geslin2021microelasticityII}. The same is then true for the pinning displacement and the corresponding correlation function value. For 60$^{\circ}$ dislocations (which are the partials of an edge dislocation in the fcc lattice) we find, using Eq. (6) with the correlation functions of \cite{geslin2021microelasticityII}, the values $w_{\rm p} =1.795 a, \Phi_{\perp}(w_{\rm p})= 0.273$. 

We can then write with Eqs. (4), (7) and (9) the length dependent pinning stress $\tau_L$ as 
\begin{equation}
	\tau_L=\left\{\begin{array}{l}\displaystyle
		\tau_{\rm c} \sqrt{\frac{L_{\rm p}}{L}} \left(\frac{1}{3 \Phi_{\perp}(L_{\rm p})}\right)\left(\frac{2\int_0^{w_L}\Phi_{\perp}(y)\dy}{w_L}-\Phi_{\perp}(w_L)\right)\;,\quad L\le L_{\rm p}\\
		\tau_{\rm c}\;,\quad L> L_{\rm p}.	
	\end{array}\right.
\end{equation}
For small $L$ values, we find an inverse square root relationship between $L$ and $\tau_L$. By fitting Eq. (27) to scale dependent flow stress data we can directly establish the pinning length $L_{\rm p}$ and critical stress $\tau_{\rm c}$.

\section{Comparison with atomistic and DDD simulation data}

To validate the complex behavior discussed in the previous section, atomistic simulations were conducted for dislocations of various orientation in an equiatomic FeNiCr alloy.  Details of the simulation method are presented in Appendix D. Bulk material parameters of the simulated alloy are compiled in Table 1 together with derivative parameters such as the SF stress $\tau_{\Gamma}$ and the theoretical SF width for screw and edge dislocations.
\begin{table}[tbh]
	\centering
\begin{tabular}{|l|l|}
	\hline 
    Parameter	& \\
	\hline
	Lattice Parameter [\AA] & 3.52 \\
	$C_{11}$[GPa] & 242.9 \\
	$C_{12}$[GPa] & 157.1 \\
	$C_{44}$[GPa] & 135.0 \\
	Stacking Fault Energy [mJ/m$^2$] & 60.5 \\
	SF stress $\tau_{\Gamma}$ [MPa] & 486 \\
	theoretical SF width, edge dislocation $d_0(\theta = \pi/2)$ [\AA] & 60.5 \\
	theoretical SF width, screw dislocation $d_0(\theta = 0)$ [\AA] & 35 \\
	\hline
\end{tabular}\\[6pt]
	\caption{\label{tab1} Bulk parameters of the simulated FeNiCr alloy and derivative parameters.}
\end{table}

\subsection{Metastability and hysteresis during loading/unloading}

Subsequent to creation and relaxation of a dislocation, an external shear stress is applied in the dislocation slip system, the system is again relaxed until it reaches a metastable configuration, and the new SF width is recorded. This process is repeated at increasing stresses until no stable configuration an be found, indicating loss of metastability. 
\begin{figure}[tb]
	\centering
	\hbox{}
	\includegraphics[width=0.6\textwidth]{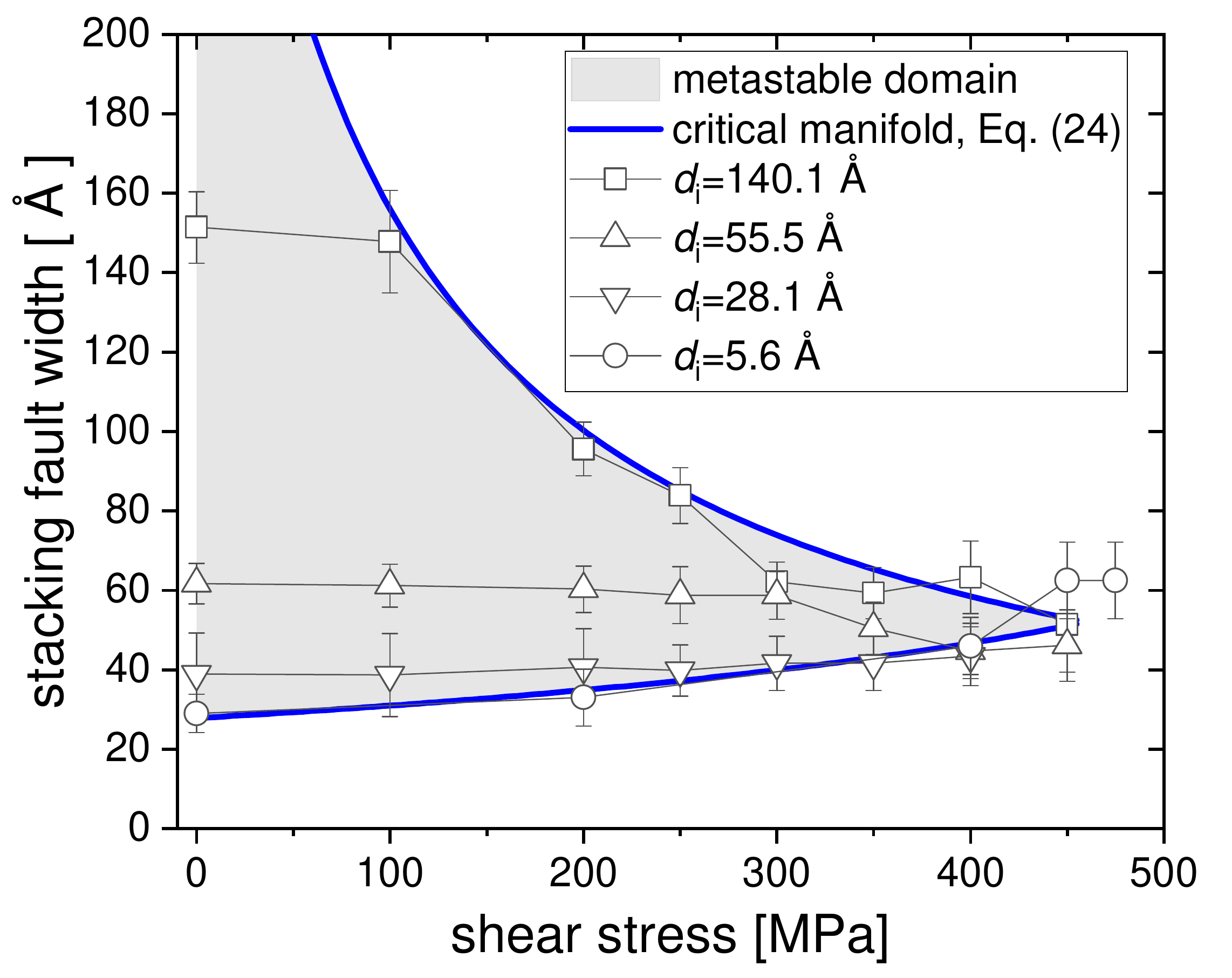}\hfill
	\caption{\label{fig:loadingpartials}
		Evolution of SF width $d$ during loading. Data points refer to atomistic simulations of the FeNiCr alloy, in these simulations the dislocations were subjected to a sequence of increasing shear stresses, relaxation was carried out after each stress step; error bars indicate the mean square variation of the SF width; blue lines: critical manifold according to Eq. (24), parameters: $\tau_{\Gamma}=529$ MPa, $\tau_{\rm ext,c}=461$ MPa, equilibrium SF width $d_0 =52$ \AA. The grey area indicates metastable states.}
\end{figure}
The observed evolution of the SF width as depicted in \figref{fig:loadingpartials} for edge dislocations in FeCrNi faithfully matches the expectations for independently pinned partials as outlined in Section 4.2. Initial SF widths that fall into the metastable domain remain unchanged until the stress reaches the critical manifold where the '=' sign holds in Eq. (24). The SF width then follows the critical manifold, i.e., it decreases with increasing stress if $d > d_0$ and it increases if $d < d_0$. Ultimately the stress reaches the critical level $\tau_{\rm ext,c}$ where the two branches merge. Fitting Eq. (24) to best represent the data yields the parameters $d_0 = 52$ \AA, $\tau_{\Gamma} = 529$ MPa, and $\tau_{\rm ext,c} = 480$ MPa. These values compare very well with the theoretical expectations (Table 1) which predict from bulk parameters $d_0 = 60.5$ \AA, $\tau_{\Gamma}= 486$ MPa, and with direct observation in the atomistic simulations which indicates loss of metastability at a stress near 480 MPa. 

\begin{figure}[tb]
	\centering
	\hbox{}
	\includegraphics[width=0.55\textwidth]{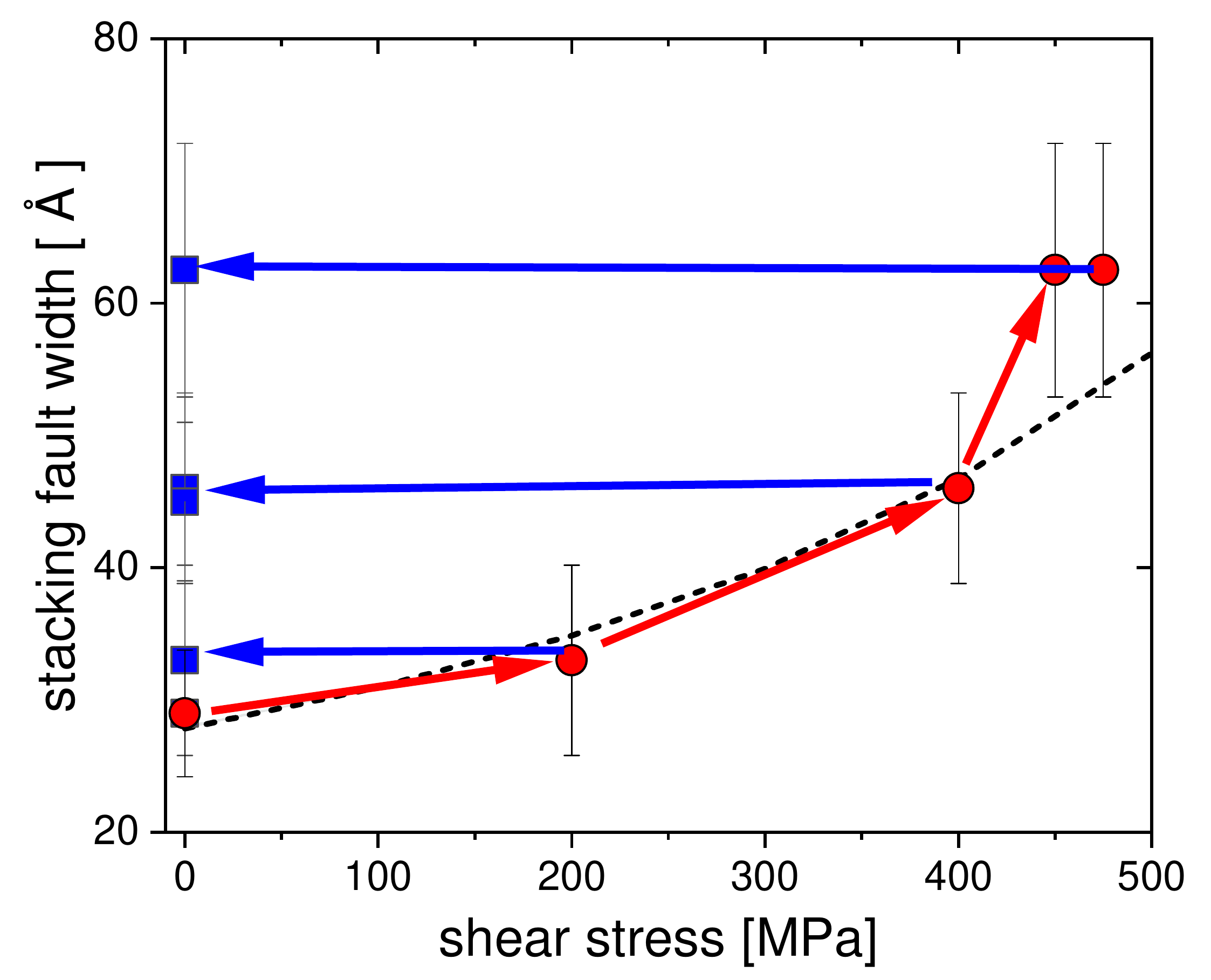}\hfill
	\caption{\label{fig:hysteresis}
		Evolution of SF width $d$ during loading-unloading cycles in atomistic simulations of FeNiCr. Red data points: SF width during loading, blue data points: SF width during unloading; red arrows indicate loading steps, blue arrows unloading steps; dashed line: lower critical manifold according to Eq. (24) with parameters as in \figref{fig:loadingpartials}.}
\end{figure}

Next we address the question of hysteresis. To this end, for an initial stacking fault width that lies on the lower critical manifold of \figref{fig:loadingpartials}, a sequence of loading steps was carried out as described above. However, after each loading step the thus reached configuration was duplicated and the duplicate unloaded to zero applied stress and then relaxed, as illustrated by the arrows in \figref{fig:hysteresis}. It can be seen that, upon unloading from the critical manifold, the SF width remains unchanged, demonstrating hysteresis. 

\subsection{Pinning of constrained dislocations}

Atomistic simulations offer a convenient opportunity to study constrained pinning. By simply imposing periodic boundary conditions in line-parallel direction, an upper limit to the length $L$ of a coherently displacing segment is imposed without compromising the elastic interactions by introduction of free surfaces, and without adding additional interactions with constraining particles or interfaces. Thus, over-constrained atomistic simulations with small periodicity length in line parallel direction offer a nice opportunity to directly probe the theoretical relations formulated in Section 4.3. 

\begin{figure}[tbh]
	\centering
	\hbox{}
	\includegraphics[width=0.6\textwidth]{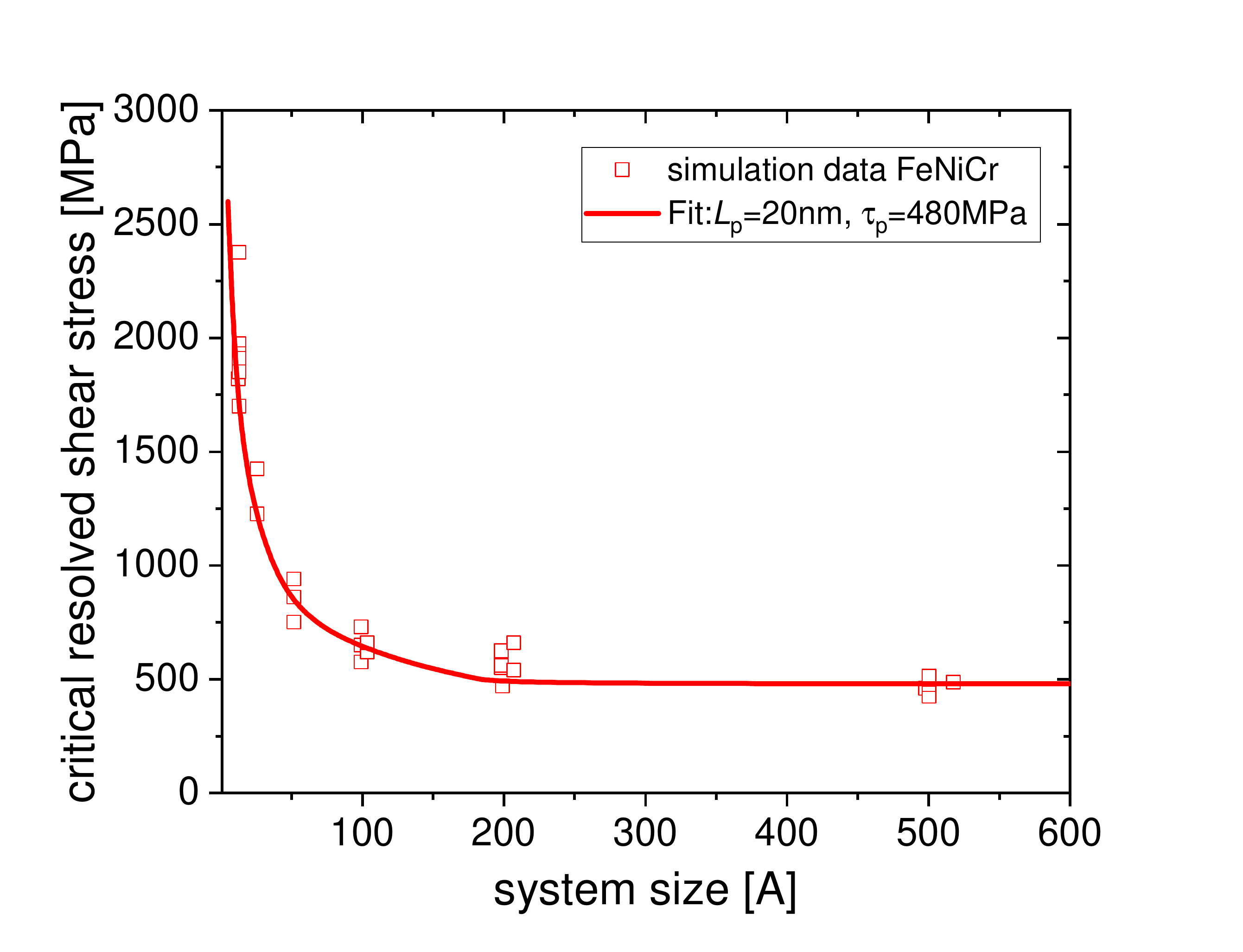}\hfill
	\caption{\label{fig:constrainedpinning}
		Pinning stress for a constrained system according to atomistic simulations of edge dislocations with periodic boundary conditions in dislocation line and glide direction, dependence of critical stress for dislocation motion on system size in line-parallel direction;  full line: Fit using Eq. (21).}
\end{figure}

We illustrate this for quasistatic simulations of constrained edge dislocations in  the FeNiCr  alloy systems. Results are shown in \figref{fig:constrainedpinning} together with fits obtained from Eqs. (26,27) where the only fit parameters used are the critical pinning stress $\tau_{\rm c}$ and the pinning length $L_{\rm p}$. The obvious qualitative observation is a very significant increase of the pinning stress (here pragmatically defined as the stress required to move the dislocation across the simulation cell) below a constraint length of about 20 nm. For constraint lengths as low as 1 nm, the pinning stress may increase up to five-fold. This observation indicates that nanometre-scale geometrical constraints to dislocation motion can significantly increase the strengthening action of solutes. Our fit of the theoretical relationship to the simulation data allow us to put this assertion on a quantitative footing. 
Together with the simulation data, \figref{fig:constrainedpinning} includes a fit based on Eqs (26,27) with the correlation functions given in the Appendix. The resulting fit parameters for FeNiCr are $L_{\rm p} = 20$ nm, $\tau_{\rm p} = 480$ MPa. These parameters provide a good match between theoretical curve and simulation data (red data points and red curve in \figref{fig:constrainedpinning}). In particular, we note that the fitted value of the unconstrained pinning stress agrees very well with the fit parameters shown in \figref{fig:loadingpartials}. The latter indicate a pinning stress of 465 MPa, and  direct observations in our simulations  indicate a depinning stress slighly above 480 MPa. The value $\tau_{\rm p} = 480$ MPa obtained from fitting constrained flow stress data to theoretical relations is in excellent agreement with these independent reference data.  

\section{Comparison with DDD simulations}

To verify whether the phenomenology of the  findings from atomistic simulations can be reproduced in mesoscopic simulations, discrete (partial) dislocation dynamics simulations were carried out using the methodology described in Ref. \cite{zhai2019properties}. Details regarding the simulation method are given in Appendix D2. Such mesoscopic simulations have the advantage that the mesoscale parameters can be varied without regard to actual material realizations. Thus, the ratio between the SF stress $\tau_{\Gamma}$ and the pinning stress $\tau_{\rm c}$ can be set to an arbitrary value by independently tuning the strength of the pinning field and the SF energy -- something that is manifestly impossible in atomistic simulations where both quantities emerge from the same atomic-scale interactions. Here we exploit this possibility for probing both the validity and the limits of the independent-pinning assumption. 

\begin{figure}[tb]
	\centering
	\hbox{}
	\includegraphics[width=0.9\textwidth]{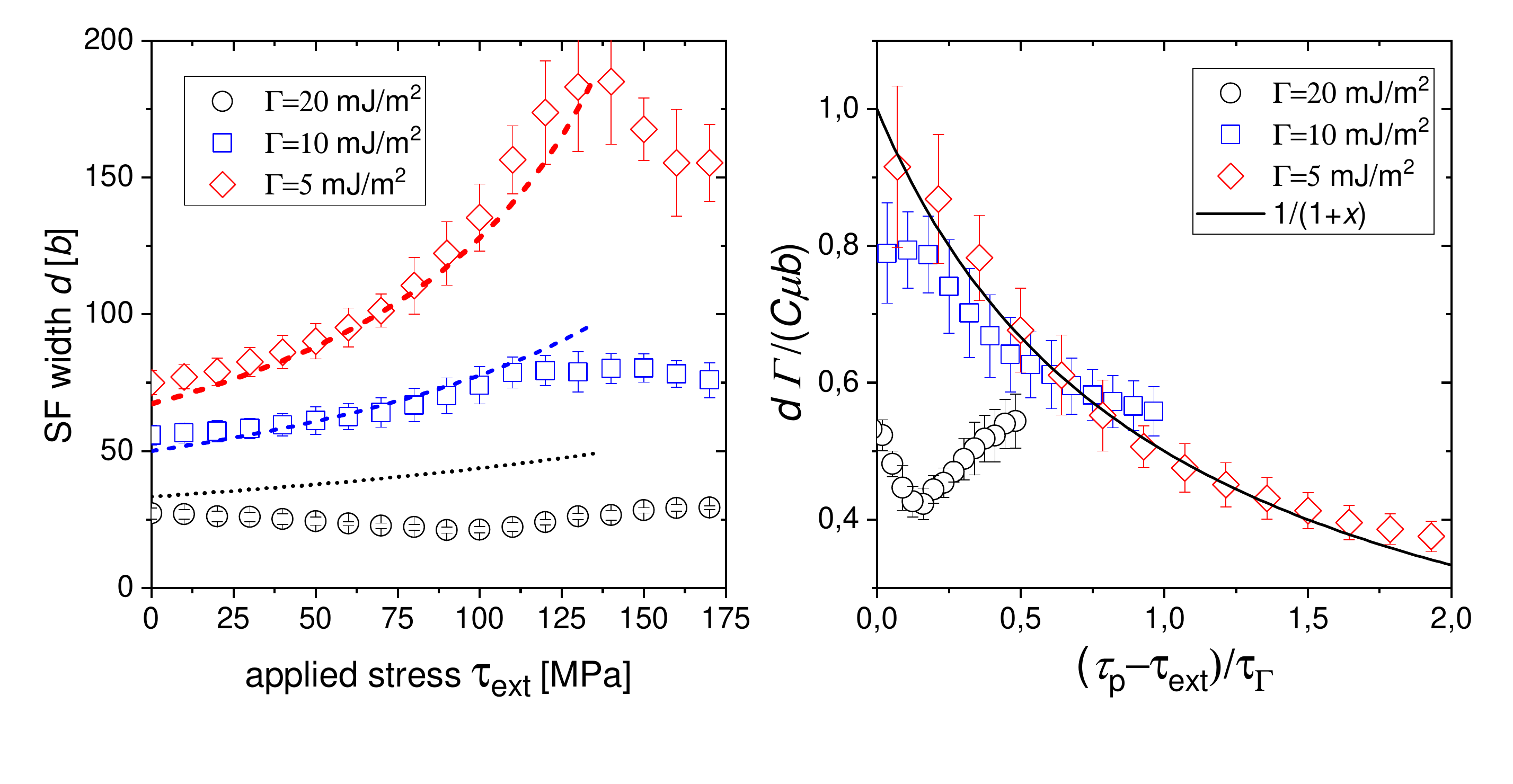}\hfill
	\caption{\label{fig:loadingpartialsDDD}
		Left: Evolution of SF width $d$ during loading in D(P)DD simulations with varying SF energies as given in the legend; the dashed lines represent fits according to Eq. (23) with pinning stress $\tau_{\rm c} = 137$ MPa, SF stress $\tau_{\Gamma} = 71$ MPa (red dashed line) and $\tau_{\rm \Gamma} = 142$ MPa (blue dashed line), the SF equilibrium width $d_0$ was evaluated according to Eq. (11) with $C_{\Gamma} = 0,109$. Right: scaled curves according to Eq. (28).}
\end{figure}

Figure \ref{fig:loadingpartialsDDD} (left) shows stress dependent SF widths determined from DDD simulations of edge dislocations conducted for material parameters of Cu. In these simulations, the amplitude of the pinning field was set to $\langle \tau^2 \rangle = ({\rm 200 MPa})^2$, while the SF energy and thus the equilibrium SF width was varied by a factor of 4. Up to a pinning stress of about 130 MPa which approximately matches the fitted pinning stress of 135 MPa, the data for the SF energy values of 5 mJ/m$^2$ and  10 mJ/m$^2$ are well represented by Eq. (23) if we note that the SF stress is proportional to the SF energy and use Eq. (11) to estimate the equilibrium SF width. The fitted value $C_{\Gamma}=0.11$ is comparable to the theoretical estimate for an edge dislocation, Eq. (12) which with $\nu = 0.3$ gives $C_{\Gamma}=0.131$. 

Moreover, validity of Eq. (23) in conjunction with Eq. (11) and the definition of the SF stress implies that a plot of $d/d_0$ vs $x = (\tau_{\rm c}-\taue)/\tau_{\Gamma}$ leads to a curve that should be independent of SF energy {\em if} the pinning stress is SF energy independent - which is a corollary to the assumption of independently pinned partials. The ensuing curve has the form
\begin{equation}
	d/d_0 = \frac{1}{1+x}\;,\; x = \frac{\tau_{\rm c}-\taue}{\tau_{\Gamma}}.
\end{equation}
It can be seen in \figref{fig:loadingpartialsDDD} (left) that this behavior is well fulfilled for the SF energy values of 5 mJ/m$^2$ and  10 mJ/m$^2$ but not for a SF energy of 20 mJ/m$^2$ where Eqs. (23) and (28) predict a behavior that is both quantitatively and qualitatively at variance with the simulation data. 

To understand the reason for the discrepancy we note that the assumption of independently pinned partials is strictly valid if the interaction of partials is negligible as compared to their interaction with the pinning field, i.e., $\tau_{\Gamma}/\tau_{\rm c} \ll 1$. If we now evaluate the ratio of the fitted SF stress and pinning stress, then we find values of $\tau_{\Gamma}/\tau_{\rm c} = 0.53$ and $\tau_{\Gamma}/\tau_{\rm c} = 1.06$
for the SF energy values of 5 mJ/m$^2$ and  10 mJ/m$^2$, respectively. The latter value compares well with the data in Figure 2 where $\tau_{\Gamma}/\tau_{\rm c} = 1.11$. The dislocation dynamics simulations with SF energy of 20 mJ/m$^2$, on the other hand, lead to a ratio $\tau_{\Gamma}/\tau_{\rm c} = 2.12$ which is evidently beyond the range of the independent pinning assumption. 

\begin{figure}[tb]
	\centering
	\hbox{}
	\includegraphics[width=0.55\textwidth]{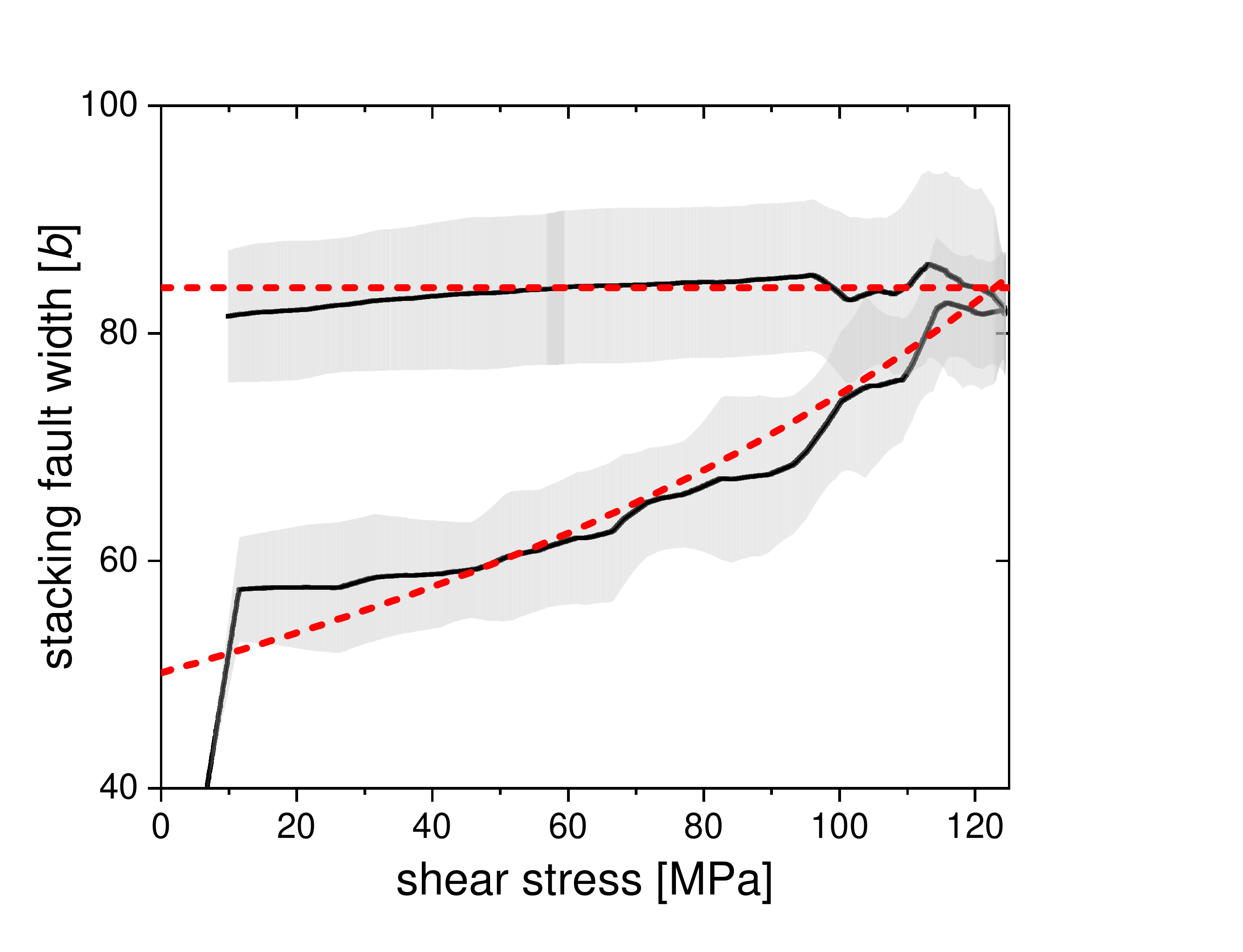}\hfill
	\caption{\label{fig:hysteresisDDD}
		Evolution of SF width $d$ during an loading-unloading cycle in a D(P)DD simulation, parameters as for blue data points in  \figref{fig:loadingpartialsDDD}, the grey shaded area denotes the error bar defined by the variance of 20 simulations with different realizations of the random pinning field, the dashed lines represent the expectation according to Eq. (23) with parameters as in \figref{fig:loadingpartialsDDD}.}
\end{figure}

If we stay within the limits of this assumption, then we observe in the D(P)DD simulations a loading-unloading hysteresis that closely matches the findings in the atomistic simulations, as illustrated in \figref{fig:hysteresisDDD} for a loading-unloading simulation conducted with a SF energy of 10 mJ/m$^2$.

\section{Discussion and Conclusions}

Our analysis reveals a complex picture of the pinning of dislocations that are dissociated into weakly coupled partials. In the classical picture of elastic manifold depinning, Middleton's 'no passing' theorem \cite{middleton1992asymptotic} ensures the existence of a unique sequence of metastable states that are crossed by an elastic manifold as the load is increased towards the depinning threshold. Here, instead, the ability of extended dislocations to adjust the stacking fault width between two partials provides an internal degree of freedom that gives, in turn, rise to a multitude of metastable states and consequently to a rich and complex behavior which includes metastability, hysteresis, as well as dynamic phenomena which include the dissociation of dislocations under load that are thermodynamically stable from the viewpoint of free energy minimization, and the dynamic stabilization of extended dislocations that are thermodynamically unstable due to a negative SF energy. 

From a practical viewpoint, the complex superposition of the mutual interactions of partial dislocations on the one hand, and their interactions with randomly distributed solutes that give rise to substantial pinning stresses on the other hand, makes it difficult if not impossible to deduce important parameters such as the SF energy straightforwardly from observations of the SF width of extended dislocations. 
This is of particular relevance regarding the current debate about the possibility of negative stacking fault energies in concentrated solid solution alloys, see e.g.,\cite{shih2021stacking,sun2021can,werner2021experimental}. 

Adding an additional level of complexity, our analysis of constrained pinning has revealed that the flow stress enhancement due to solute effects is increased if the dislocation is constrained. This implies that different hardening mechanisms are not merely additive, but may behave in a super-additive manner such that, for instance, the presence of nanoscale precipitates or twins may boost the strengthening action of solutes -- a synergetic effect that might render simple addition rules (like additivity of the stress contributions or additivity of the stress squares), which are commonly used in the literature, spurious once the length scales associated with such hardening mechanisms become comparable to the dislocation pinning length. 

In view of dislocation simulations, the present study points to the importance of moving beyond DDD towards a partial dislocation dynamics (DPDD) in situations where solute pinning effects are prominent. According to our findings the ratio of the SF stress $\tau_{\Gamma}$ and the pinning stress $\tau_{\rm c}$ provides a simple qualitative criterion, $\tau_{\Gamma}/\tau_{\rm c} \approx 1$, for delineating the regime independent pinning $\tau_{\Gamma}/\tau_{\rm c} \le$ where dislocations can be envisaged as near-independent partials.

\section*{Declarations}
\subsection*{Competing interests}
  The authors declare that they have no competing interests.
\subsection*{Author's contributions}
A.V. implemented and performed the atomistic simulations with support from E.B. D.W. implemented and performed the DDD simulations. S.N checked calculations and analyzed data using scaling arguments and theoretical relations formulated by M.Z. M.Z. drafted the manuscript which was edited and approved jointly by all authors. 
\subsection*{Funding}
M.Z. and E.B. acknowledge funding by DFG under Grants no. Za 171/8-1 and Bi-1453/2-1. 
\subsection*{Acknowledgements}
E.B. would like to thank Fritz Körmann for valuable scientific discussion.
The atomistic simulations were performed on computer resources provided by the Erlangen Regional Computing Center (RRZE).
\subsection*{Availability of data and material}
Not applicable

\bibliographystyle{bmc-mathphys} 
\bibliography{references}   

\newpage

\appendix
\setcounter{equation}{0}
\renewcommand{\theequation}{\Alph{section}.\arabic{equation}}
\section{Correlation functions and their integrals}

Correlation functions for shear stresses created by a random arrangement of dilatation and/or compression centers were evaluated in Ref. \cite{geslin2021microelasticityII}. Here we give these expressions together with the corresponding integrals which are needed for evaluation of pinning parameters. 

The correlation functions of shear stresses in a slip plane are anisotropic with respect to the direction of shear. Hence, one needs to distinguish shear stress correlations in direction of shear, described by a 'longitudinal' correlation function $\Psi_{\rm L}$, and stress correlations in perpendicular direction described by a 'transverse' correlation function $\Psi_{\rm T}$. These functions are given by \cite{geslin2021microelasticityII}
\begin{eqnarray}
	\Psi_{\rm L}(u)&=&-\frac{30}{u^3} \left[\sqrt{\pi}\left(
	1-\frac{12}{u^2}\right)\erf\left(\frac{u}{2}\right)+\frac{1}{u}
	\left(12 + u^2\right) \exp\left(-\frac{u^2}{4} \right)\right],
	\nonumber\\
	\Psi_{\rm T}(u) &=& \frac{15}{u^3} \left[\sqrt{\pi}\left(
	1-\frac{6}{u^2}\right)\erf\left(\frac{u}{2}\right)+\frac{6}{u}
	\exp\left(-\frac{u^2}{4}\right)	\right],
\end{eqnarray}
where $u$ is the distance from the origin in the respective direction. For evaluating pinning parameters, also the integrals of these functions are needed which are given by 
\begin{eqnarray}
	\int_0^U\Psi_{\rm L}(u){\rm d}u &=&\frac{15}{U^4} \left[\sqrt{\pi}\left(
	U^2-6\right)\erf\left(\frac{U}{2}\right)+  6\exp\left(-\frac{U^2}{4} \right)\right],
	\nonumber\\
	\int_0^U\Psi_{\rm T}(u){\rm d}u &=& \frac{15}{8U^4} \left[\sqrt{\pi}\left(
	U^4-4U^2+12\right)\erf\left(\frac{U}{2}\right)+(2U^3-12U)
	\exp\left(-\frac{U^2}{4}\right)	\right],\nonumber\\
\end{eqnarray}
From these expressions,  the line-parallel and line-perpendicular correlation functions for a dislocation that forms an angle $\vartheta$ with respect to the direction of shear are obtained by superposition according to
\begin{eqnarray}
	\Phi_{\parallel}(u) &=& \Psi_{\rm L}(u) \cos^2\vartheta +  \Psi_{\rm T}(u) \sin^2\vartheta,\nonumber\\
	\Phi_{\perp}(u) &=& \Psi_{\rm L}(u) \sin^2\vartheta +  \Psi_{\rm T}(u) \cos^2\vartheta.
\end{eqnarray}
The respective integrals are calculated in an analogous manner. In particular, the correlation length (defined as the indefinite integral of the line-parallel correlation function) derives as 
\begin{equation}
	\xi(\theta) = \frac{15}{4} \sqrt{\pi} \sin^2 \vartheta.
\end{equation}

\section{Interaction of fluctuating partial dislocations}

The general stress field generated by a dislocation loop parameterized by $\Br' = \Br'(s)$ is given by the Peach-Koehler formula
\begin{eqnarray}
	\sigma_{ab}(\Br) &=& \frac{\mu}{8\pi}\oint \partial_i\Delta R b_k [\epsilon_{ika} \ds_b + \epsilon_{iba} \ds_a] \nonumber\\
	&+& \frac{\mu}{4\pi (1-\nu)}\oint b_k \epsilon_{ikl}[\partial_i\partial_a\partial_b R - \delta_{ab} \partial_i \Delta R] \dx_l
\end{eqnarray}
where $R = \sqrt{(\Br-\Br').(\Br-\Br')}$ and $\ds_k = \Be_k.({\rm d}\Br'/
\ds)$. 

We consider the stress produced by a partial dislocation 1 parameterized by $(x_1, d + y_1(x_1))$. The dislocation is contained in the $xy$ plane and the only nonvanishing Burgers vector components are $b_x$ and $b_y$. The vector line element has the components $\ds_x = \dx_1$ and $\ds_y = \partial_{x_1}y_1 \dx_1$. 

The glide component of the Peach-Koehler force acting on a second dislocation with Burgers vector components $b'_x$ and $b'_y$ that is lying along the $x$ axis with fluctuations $y_2(x_2)$ is given by $f_{\rm PK,2} = b'x \sigma_{xz} + b'y \sigma_{yz}$. We thus need to evaluate the $xz$ and $yz$ components of the stress field of the first dislocation. We evaluate these components as the sum of 
the stress field of a straight dislocation at $y = d$ (the 'average dislocation') plus a fluctuating loop that is composed of the actual dislocation and the negative average dislocation, see Figure A1. In evaluating the integrals, we use the following relations:
\begin{eqnarray}
	\int_{-\infty}^{\infty} \partial_j \partial_k \partial_l R (\dy_1/\dx_1) \dx_1 &=&
	- \int_{-\infty}^{\infty} y \partial_y  \partial_j \partial_k \partial_l R\dx\nonumber\\
	\partial_i \partial_j \partial_k [R(x,d_0+y,0) - R(x,d_0,0)] &\approx& 
	y \partial_y \partial_i \partial_j \partial_k R(x,d_0+y,0)|_{y=0} 
\end{eqnarray}
With these relations we find for the relevant stress components the following results: 
\begin{eqnarray}
	\sigma_{xz}(x,y) &=& \sigma_{xz}^0 + \frac{\mu b_x}{8\pi} \int_{-\infty}^{\infty}
	\left[\partial_y^2 \Delta R_d + \frac{2}{1-\nu} \partial_x^2 \partial_z^2 R_d 
	\right]y_1(x)\dx\\
	\sigma_{yz} &=& \sigma_{yz}^0 + \frac{\mu b_y}{8\pi} \int_{-\infty}^{\infty}
	\left[\partial_x^2 \Delta R_d + \frac{2}{1-\nu} \partial_y^2 \partial_z^2 R_d 
	\right]y_1(x)\dx
\end{eqnarray}
where $R_d = \sqrt{(x_2-x_1)^2 + (d + y_1(x_1)-y_2(x_2))^2 + z^2}$ and the derivatives are evaluated at the point $y_1=y_2=z=0$. We discuss two asymptotic cases: 

\begin{enumerate}

\item
The characteristic length scale $L$ of variations of $y(x)$ is much larger than $d$. In this case the relations simplify to
\begin{eqnarray}
	\sigma_{xz}(x_2,y_2) &=& \sigma_{xz}^0(d) \left(1 - \frac{y_1(x)}{d}\right) + {\cal O}(d/L),\\
	\sigma_{yz}(x,y) &=& \sigma_{yz}^0(d) \left(1 - \frac{y_1(x)}{d}\right) + {\cal O}(d/L).
\end{eqnarray}
The resulting Peach-Koehler force on the second partial can, up to terms of order $d/L$, be expressed as
\begin{eqnarray}
	f_2 &=& b'_x \sigma_{xz}^0(d)  \left(1 + \frac{y_2(x)}{d} - \frac{y_1(x)}{d}\right) + b'_y \sigma_{yz}^0(d_0)  \left(1 + \frac{y_2(x)}{d} - \frac{y_1(x)}{d}\right)\nonumber\\
	& =& \frac{C_{\gamma}\mu b_{\rm p}^2}{d}\left(1 + \frac{y_2(x) - y_1(x)}{d}\right). 
\end{eqnarray}
An analogous expression, with signs of $y_1$ and $y_2$ reversed, derives for the first dislocation. Adding the force due to the stacking fault, we find for the total interaction force 
\begin{equation}
	f_{1,\Gamma} = - f_{2,\Gamma} = \frac{C_{\gamma}\mu b_{\rm p}^2}{d_0}\left(\frac{d_0}{d} - 1 + \frac{y_2(x) - y_1(x)}{d}\right) 
\end{equation}
The energy contribution associated with the interaction takes the simple form
\begin{equation}
E_{\Gamma} =  C_{\gamma}\mu b_{\rm p}^2 \left(\left[\frac{1}{d} - \frac{1}{d_0}\right](y_1-y_2) + \frac{(y_1-y_2)^2}{2d^2}\right)
\end{equation}

\item
The characteristic length scale $L$ of variations of $y(x)$ is much smaller than $d_0$. In this case the stress field associated with a small localized perturbation has the character of a dipole field:
\begin{eqnarray}
	\sigma_{xz}(x,y) &=& \sigma_{xz}^0 \left(1 + {\cal O}(Ly_1/d^2)\right),\\
	\sigma_{yz}(x,y) &=& \sigma_{yz}^0 \left(1 + {\cal O}(Ly_1/d^2)\right). 
\end{eqnarray}
Neglecting the dipolar contribution due to the shape fluctuations, one finds that the main contribution to the interaction is just the interaction of each partial with the average opposite partial dislocations, i.e.,
\begin{equation}
	f_{1,\Gamma} = \frac{C_{\gamma}\mu b_{\rm p}^2}{d_0}\left(\frac{d_0}{d} -1 - \frac{y_1}{d}\right)\;,\;
	f_{2,\Gamma} = \frac{C_{\gamma}\mu b_{\rm p}^2}{d_0}\left(-\frac{d_0}{d} + 1 - \frac{y_2}{d}\right)\;,\;
\end{equation}
This expression can be simplified by noting that, in the limit $d \gg L$ of weakly interacting partials, the terms proportional to $y_{1,2}$ are small in compraison with the respective line-tension terms such that we may approximate
in leading order
\begin{equation}
	f_{1,\Gamma} = - f_{2,\Gamma}\frac{C_{\gamma}\mu b_{\rm p}^2}{d_0}\left(\frac{d_0}{d} -1\right)\;.
\end{equation}
\end{enumerate}

\section{Average work of pinning}

When a segment of a perfect dislocation of length $L$ displaces over a distance $w$, the associated energy change per unit length is
\begin{equation} 
	\Delta E(L,w) = E_{\rm LT} - \frac{1}{L}\int_{-L/2}^{L/2} \int_0^w \tau(x,y) \dy \dx 
\end{equation}
where we have assumed without loss of generality that the centerpoint of the segments is initially located in the origin. We require the displacement to be in the direction of the mean stress acting on the segment while it is in its initial straight configuration, hence, the energy change is always negative. To evaluate the average magnitude of the energy change, we need to perform the average over $\tau(x,y)$ conditional on its value at $y=0$. Note that a simple spatial average will not do, since it gives displacements of the segment in the direction opposite to the stress acting on it (which do not occur) equal weight as displacements in the direction of the acting stress, with the erroneous result that the work done by the fluctuating stress in displacing the dislocation is zero on average.  

Taking the average of $\tau(x,y)$ conditional on the field value at $(x,0)$, we use $\langle \tau(x,y)|\tau(x,0)\rangle = \tau(x,0)\Psi_{\perp}(y)$. This gives
\begin{equation} 
	\Delta E(L,w) = C_{\rm LT}\mu b^2\frac{w^2}{L^2} - \frac{1}{L}\int_{-L/2}^{L/2} \tau(x,0) \dx \int_0^w \Psi_{\perp}(y) \dy 
\end{equation}
Setting the variation of this expression with respect to $w$ to zero gives 
\begin{equation} 
	2C_{\rm LT}\mu b^2\frac{w}{L^2} = \frac{1}{L}\int_{-L/2}^{L/2} \tau(x,0) \dx \Psi_{\perp}(w).   
\end{equation}
Note that the sign of $w$ here automatically equals the sign of the stress average. We now square both sides, average over the random field, and take the square root. This gives
\begin{equation} 
	2 C_{\rm LT}\mu b^2\frac{w}{L^2} = \left(\langle \tau^2 \rangle \frac{1}{L}\int_{-L/2}^{L/2} \Phi_{\parallel}(x) \dx\right)^{1/2} \Psi_{\perp}(w) \approx \left(\langle \tau^2 \rangle \frac{\xi}{L}\right)^{1/2} \Psi_{\perp}(w)   
\end{equation}
In the last step we have used that, for large $L \gg \xi$, the integral can be replaced by the definite integral over the real axis, $\xi = \int \Psi_{\parallel} \dx$.

In the next step, we consider the variation of the energy functional with respect to the length $L$ of the displaced segment. Using the same steps and making the same approximations as above we find
\begin{equation} 
	2 C_{\rm LT}\mu b^2\frac{w^2}{L^3} = \frac{1}{2}\left(\langle \tau^2 \rangle \frac{\xi}{L^3}\right)^{1/2} \int_0^w\Psi_{\perp}(x)\dx  
\end{equation}
Multiplication with $L/w$ and elimination of the left-hand side gives the equation for the characteristic pinning displacement $w_{\rm p}$ which optimizes the energy gain:
\begin{equation} 
	2 \Psi_{\perp}(w_{\rm p}) = \int_0^{w_{\rm p}} \Psi_{\perp}(x)\dx.   
\end{equation}

\section{Simulation Methods}

\subsection{Atomistic simulations}
 
The atomistic simulations were performed using the molecular dynamics code LAMMPS \cite{plimpton1995fast}. Interactions between atoms in the equiatomic, ternary solid solution FeNiCr were modelled using the embedded atom method (EAM) potential by Bonny \etal\cite{bonny2011interatomic} that was fitted to reproduce the properties of individual elements as well as the SFE and elastic properties of different alloy compositions. 
This EAM potential has been successfully used for the study of dislocations in concentrated solid solution alloys (CSSAs), e.g., in \cite{Varvenne2016_AM,osetsky2019two,dou2021interaction}.
 
The random CSSA samples were generated by randomly placing atoms of the different constituent elements on a regular fcc grid. The bulk properties, like the lattice parameter $a$ and elastic constants, were determined from a cubic sample ($\approx{} 140\times140\times140$\AA$^3$, with crystallographic axes along $[100]$, $[010]$, and $[001]$) that was relaxed under periodic boundary conditions while allowing for the adjustment of the simulation box size. 
The SFE was determined using a different setup ($\approx 100\times100\times200 \text{\AA}^3$, crystallographic axes along $[112]$, $[\bar{1}10]$, and $[\bar{1}\bar{1}1]$) by freely relaxing along $[\bar{1}\bar{1}1]$, while precluding the motion of atoms along the other directions. 
The bulk properties are reported in Tab.\ref{tab1}.
 
\begin{figure}
        \centering
        \includegraphics[width=10cm]{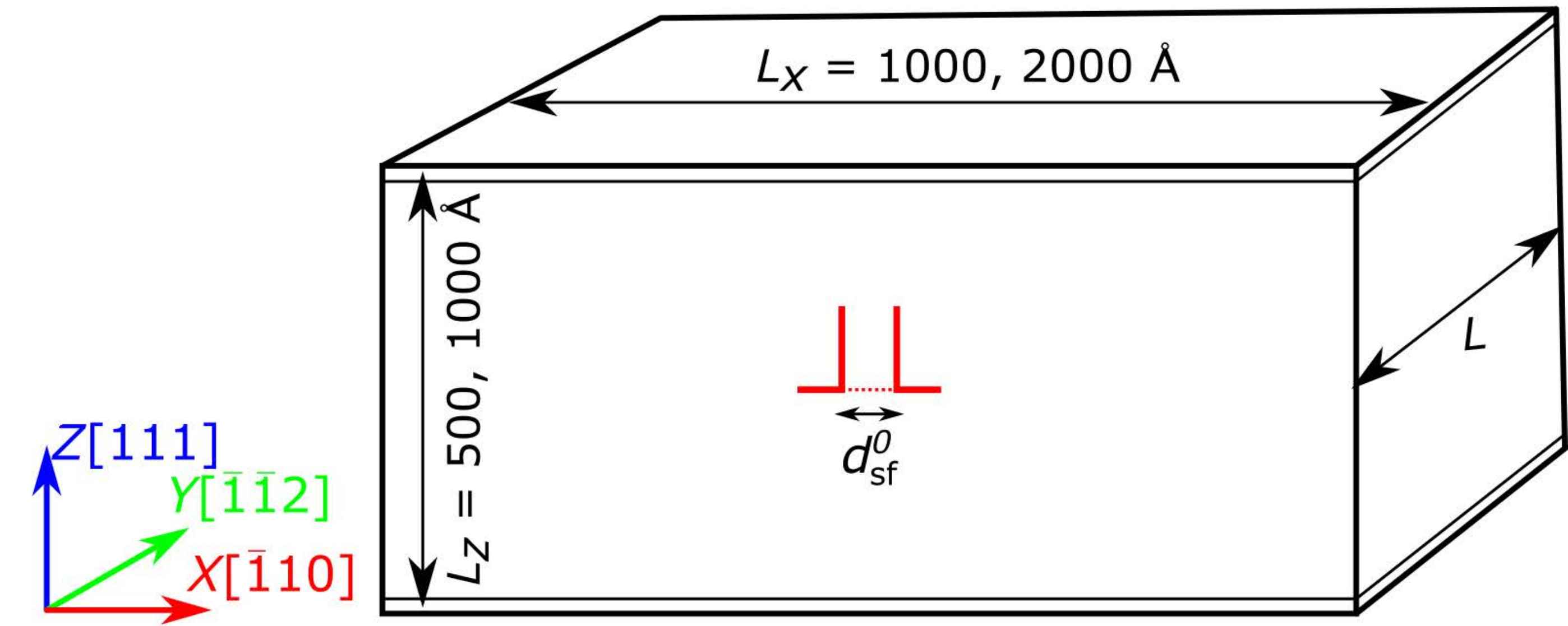}
        \caption{Simulation setup for a dislocation. An edge dislocation is shown as an example here. The simulation box is periodic along the dislocation line direction ($Y-$ axis) and along the dislocation glide direction ($X-$ axis). The atoms in the boundary layer are free to move only parallel to the $XY$ plane. The dislocation character (edge, 60$^{\circ}{}$mixed, screw), the dislocation line length ($L \approx{} 10, 25, 100, 200, 500$ \AA), and the initial dislocation splitting width ($d_{\text{i}} \approx{} 5.6, 28.1, 55.5, 140.1$\AA) are the simulation parameters.}
        \label{fig:SimulationSetup}
\end{figure}
 
A slab setup as shown in \figref{fig:SimulationSetup} was used to study dislocation properties. 
The crystallographic orientation of the slab was chosen such that the glide plane normal [111] was parallel to the $Z$ axis, the (average) line direction of the dislocation was parallel to the $Y$ axis and the glide direction was parallel to the $X$ axis. 
Periodic boundary conditions were used along $X$ and $Y$, while atoms in the top and bottom layer were only allowed to move on the $xy$ plane (2D dynamic boundary conditions).
Different sized boxes were used, with the standard box dimensions of  approximately $2000 \times L \times 500 \text{\AA}^{3}$. 
The results were not affected by the used box dimensions in $X$ and $Z$.
 
Full dislocations of edge, 60$^{\circ}{}$ mixed and screw characters were inserted in the center of the simulation samples following the method described in \cite{rodney2004molecular}, which allows for an adjustable initial splitting distance, $d_i$, of the partial dislocations.
For each combinations of dislocation line length, initial splitting distance, and dislocation character used, 20 different random atomistic configurations were generated.
 
Throughout the study, all relaxations in the static and quasistatic calculations were performed using an optimized implementation of the fast inertial relaxation engine (FIRE) algorithm to reach the equilibrium state \cite{guenole2020assessment}. 
A configuration was considered to be relaxed if the force-norm fell below the threshold value of $10^{-4}$eV/\AA. 
 
The simulation methodology to study dislocations under an externally applied shear stress follows \cite{bitzek2004atomistic}. First the relaxed configuration of a dislocation was pre-sheared according to the desired resolved shear stress. To this starting configuration forces corresponding to the desired resolved shear stress were applied to atoms in the top- and bottom- most layers parallel to the glide plane. 
Once an energy minimized configuration was found at a given stress, this was used as the initial configuration for a next relaxation step at a higher stress. 
The applied shear stress was ramped up until the dislocation could no longer find a stable energy minimized configuration, \textit{i.e.}, the dislocation had traversed the length of the simulation box at least once.
 
The adaptive common neighbor analysis (a-CNA)  as implemented in OVITO \cite{stukowski2010extracting} was used to identify the partial dislocation cores and the stacking fault.
The splitting distance $d$ of one configuration was determined by measuring the maximum width of the stacking fault area $\Delta x$ in stripes of width $\Delta y=\sqrt{6}*a/2, \sqrt{6}*a/2$, and $a/\sqrt{2}$ for edge, 60$^{\circ}$ mixed, and screw dislocations along the $Y$ axis. The average splitting distance is an average of the splitting distance along the dislocation line. The error bars in Figs.\ref{fig:loadingpartials} and \ref{fig:hysteresis} correspond to two times the standard deviation around the mean.

\subsection{Partial dislocation dynamics}

D(P)DD simulations were performed based on the PARADIS nodal dislocation dynamics code which was adapted to account for the force between two partial dislocations exerted by a stacking fault, and for the presence of a random pinning field. The SF force was represented as a constant force per unit length of magnitude $\Gamma$. This force acts perpendicularly on each line segment and is distributed evenly on the nodal endpoints. The SF force is taken positive for the trailing and negative for the leading partial. 

The pinning field was modelled as described in Ref. \cite{zhai2019properties}, as a sum of randomly located Gaussian pinning centers located on the dislocation slip plane whose width and average spacing were both taken to be $\xi = b$, mimicking atomic scale disorder. The strength of the pinning centers was adjusted to produce a mean amplitude of the resulting shear stress field on the slip plane of $\langle \tau^2 \rangle \approx (200 {\rm MPa})^2$. 

The size of the simulation box is $1000 b$ in all three spatial directions, with periodic boundary conditions. The mean dislocation segment length (nodal spacing on the partial dislocations) is equal to $b$, such as to allow the dislocation to adjust to the pinning field on all scales without imposing an {\em a priori} cutoff. The initial configuration of the dislocation represents an edge dislocation with straight partials that is threading the simulation box perpendicularly. The initial partial dislocation spacing is taken as $d_i = 40 b$, the regularization cut-off of the stress field was taken to be $a = b/2$, and the parameter controlling the core energy correction $a_0 = 0.1b$.

Material parameters used were similar to Cu ($\mu = 54.6$ MPa, $\nu$ = 0.3, $b=2.6 \AA$). The SF energy was considered an adjustable parameter that allows us to scan different pinning regimes and establish the limits of the assumption of independently pinned partials. Hence, it is understood that the DPDD simulations do not refer to any actual material.

\end{document}